\documentclass{iopconfser}

\usepackage{a4wide,amssymb,cite,graphicx}
\usepackage{epsfig}
\usepackage{hyperref}
\usepackage{tikz}
\usepackage{float}
\usepackage{graphicx}
\usepackage{subcaption}
\usepackage{ragged2e}
\usepackage{fancyhdr}
\begin{document}

\noindent Conference Proceedings for BCVSPIN 2024: Particle Physics and Cosmology in the Himalayas\\Kathmandu, Nepal, December 9-13, 2024 

\title{Simulating Negative Hydrogen ion acceleration in
LINAC-4 using Unity 3D}

\author{D.M.C.M.K. Dissanayake$^{1}$, N. Wickramage$^{1}$,   K.M. Liyanage$^{1}$,  K.A.S. Lakshan$^{1}$ }

\affil{$^1$Department of Physics, University of Ruhuna, Matara, Sri Lanka}

\email{dissanayake10749@usci.ruh.ac.lk}

\begin{abstract}
    The Linear Accelerator 4 (LINAC-4) is designed to accelerate negative hydrogen ions (H-) to high energies for
    injection into the Large Hadron Collider (LHC), where it has been supplying proton beams since 2020. LINAC-4
    accelerates negative hydrogen ions to 160 MeV. The accelerator pulses 400 microsecond bursts of negative
    hydrogen ions. A complex process is processed in this machine to achieve the required energy and simulating
    the machine is beneficial for improved understanding, visualization, educational purposes, and for training
    and skill development. Normally, high-performance, specialized software is needed to simulate the intricate
    dynamics of ion acceleration. However, our goal is to develop an interactive, real-time model that effectively
    simulates the acceleration of $H^{-}$ ions through the LINAC-4 stages by utilizing Unity 3D’s simulation and visu-
    alization capabilities. In this study, Electric fields generated by each unit of LINAC-4 are assumed as known
    wave functions, and instantaneous velocity is calculated using fundamental physics laws. It is assumed that
    a Sine wave-formed electric field is generated by the Radio-Frequency Quadrupole (RFQ) and Drift Tube
    Linacs (DTL) and a Standing wave-formed electric field is generated by Coupled-Cavity Drift Tube Linacs
    (CCDTL) and Pi-Mode Structures (PIMS). 3D structures were developed to approximate scale, and the user
    can control the frequency of the electric field to adjust the velocity, allowing the particle to reach the required
    energy level at the end of each accelerating unit. A time-based pattern is used in the Chopper line, where the
    user can change the beam-on time. Users can visualize the particle beam path and approximated graphical
    representation of the velocity variation in LINAC-4 under controllable electric fields and different time-based
    chopping patterns generated by the chopper line. This simulation creates an accessible platform for students,
    researchers, and technicians to learn about LINAC-4’s processes in an interactive manner. It also guides users
    to better understand complex acceleration physics by allowing them to visualize the ion pathways, energy
    changes, and field interactions.
\end{abstract}
\newpage
\section{Structural design and working principles of LINAC-4}

The LINAC-4 consists of six main machines, $H^-$ source, Radio Frequency Quadrupole (RFQ), Chopper Line, Drift Tube Linac(DTL), Cell Coupled Drift Tube Linac(CCDTL), and PI Mode Structures(PIMS).

\begin{center}
    \begin{tikzpicture}[node distance=1cm, every node/.style={draw, minimum width=1.5cm, minimum height=1cm, align=center}]
        \node (box1) at (0, 0) {PIMS};
    \node (box2) at (2.5, 0) {CCDTL};
    \node (box3) at (5.0, 0) {DTL};
    \node (box4) at (7.5, 0) {CHOPPER};
    \node (box5) at (10.0, 0) {RFQ};
    \node (box6) at (12.5, 0) {$H^-$ SOURCE};
    
        \draw[->,ultra thick] (box2.west)-- ++(-0.3cm,0)-- (box1.east);
        \draw[->,ultra thick] (box3.west)-- ++(-0.3cm,0)-- (box2.east);
        \draw[->,ultra thick] (box4.west) -- ++ (-0.3cm,0)-- (box3.east);
        \draw[->,ultra thick] (box5.west) -- ++ (-0.3cm,0)-- (box4.east);
        \draw[->,ultra thick] (box6.west) -- ++(-0.3cm,0)-- (box5.east);
    \node at (box5.west) [above,yshift=0.7cm] {3MeV}; 
    \node at (box3.west) [above,yshift=0.7cm] {50MeV}; 
    \node at (box2.west) [above,yshift=0.7cm] {100MeV}; 
    \node at (box1.west) [above,yshift=0.7cm] {160MeV}; 
 
    \end{tikzpicture}
\end{center}

The negative hydrogen ions are produced in the $H^-$ source and injected into the Low Energy Beam Transport, which is used to transport and match the beam from the source and match into the RFQ input parameters \cite{Linac4_Design}.
In LINAC-4 RFQ is responsible for boosting the beam up to the required energy level for injection into the DTL \cite{vretenar2008status}.
RFQ is designed to accelerate and bunch ion beams using a radio frequency electric field.
After accelerating the beam up to 3MeV, it is injected into the Chopper Line where it is chopped into pulses to get the frequency same as the PS Booster.
Next beam is injected into DTl, where the beam will be accelerated 3 to 50 MeV, using a regularly oscillating electric field. The particles are moved through a series of Drift tubes (which worked as shields during change of the electric field) and accelerate in the gaps between them.

After boosting up to 50 MeV beam is injected into CCDTL, which has 27 small DTL tanks each consisting of three accelerating gaps. 
The beam is accelerated up to 100MeV at the CCDTL and injected into PIMS, composed of seven coupled cells, which are operating at 352 MHz in $\pi$ mode.
In PIMS beam is accelerated up to 160MeV \cite{CERN_Linac4}.

\section{Simulation Model Development}

The approximate 3D model for LINAC-4 simulation  was built in Blender from scratch, according to the technical design report of LINAC-4 \cite{Linac4_Design}.

\begin{figure}[H]
    \centering
    \includegraphics[width=0.5\textwidth]{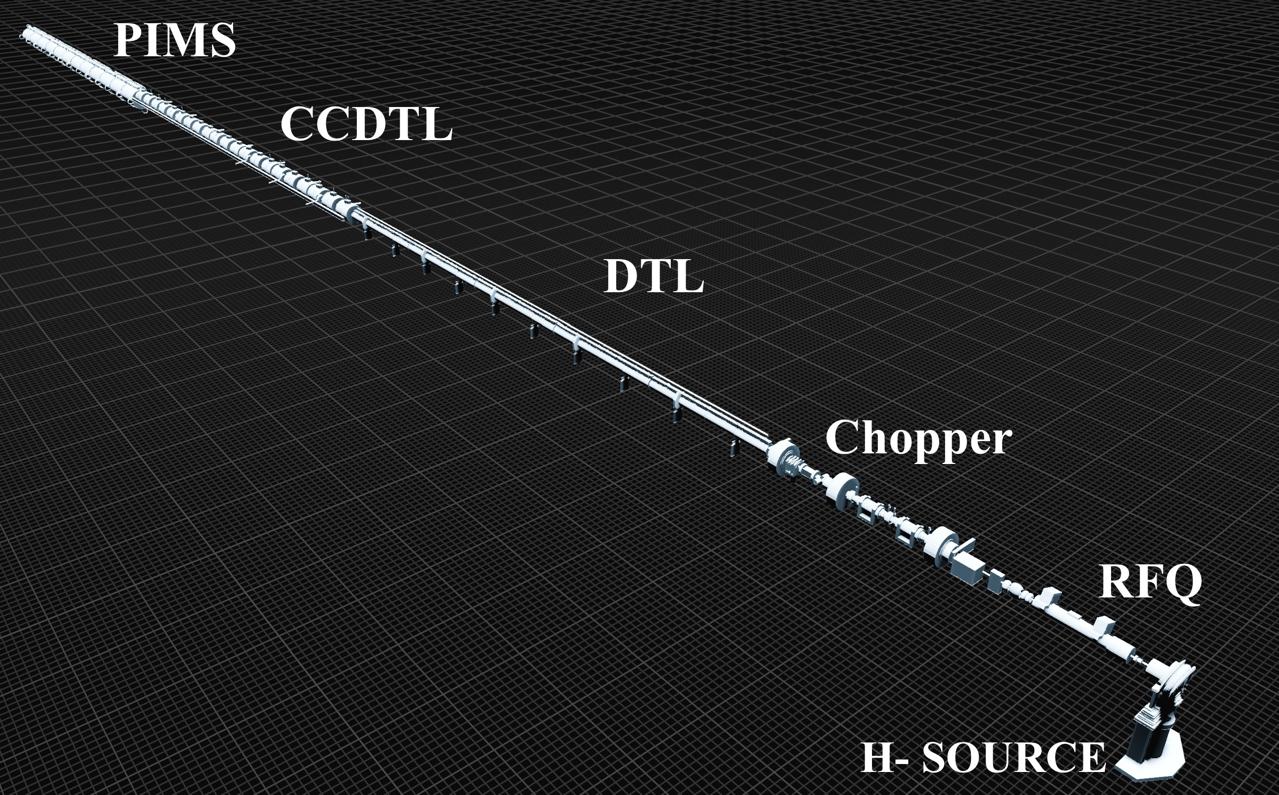}
    \caption{ 3D Model of LINAC-4 }
    \label{fig:Linac_4sub_}
\end{figure}

The algorithm for this Simulation was designed such that an electric field force was added to the rigid body of the $H^-$ particle when it reached the LINAC-4  area.
This electric field force is calculated using the following formula,

\begin{equation}
    F_E  = E * q
    \label{eq: algorithm for Electric force}
    \end{equation}

    \begin{itemize}

        \item\(E\):Instantaneous Electric field strength 
        \item\(F_E\):Electric Force 
        \item\(q\): Charge of the particle
    
    \end{itemize}

The generated electric field strength will be assigned to the E of this equation and Electric force is generated using it.
The Electric Force was generated by multiplying the Electric Force value with a unit vector in the opposite direction to the moving direction of the $H^-$ particle.

\subsection{Radio Frequency Quadrupole (RFQ) }

\raggedright
\justify
The 3D model for the RFQ was designed as the figure, \label{fig: RFQ_model}

\begin{figure}[H]
    \centering
    \begin{subfigure}{0.3\textwidth}
        \includegraphics[width=\linewidth]{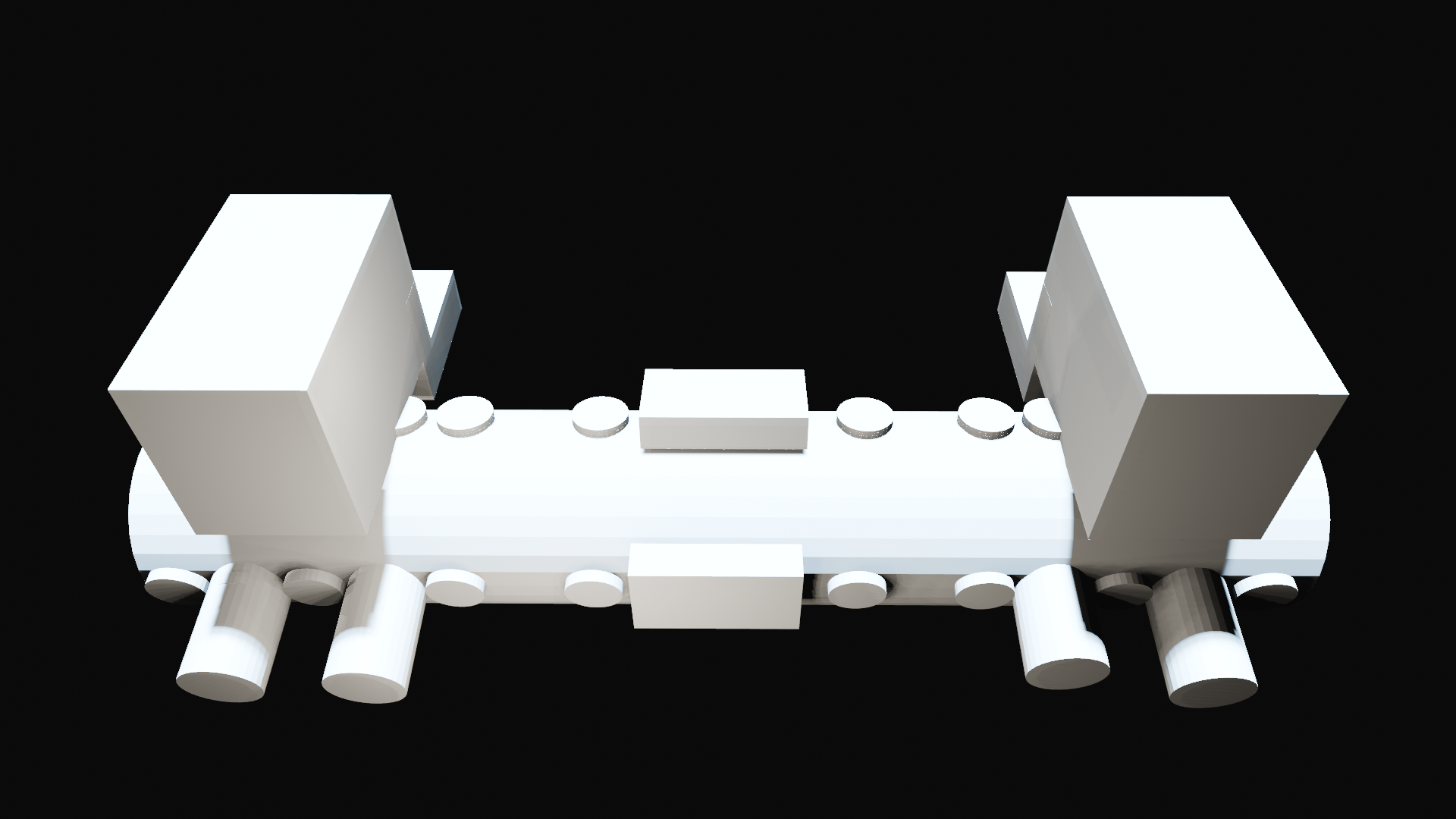}
        \caption{Front-View}
    \end{subfigure}
    \hfill
    \begin{subfigure}{0.3\textwidth}
        \includegraphics[width=\linewidth]{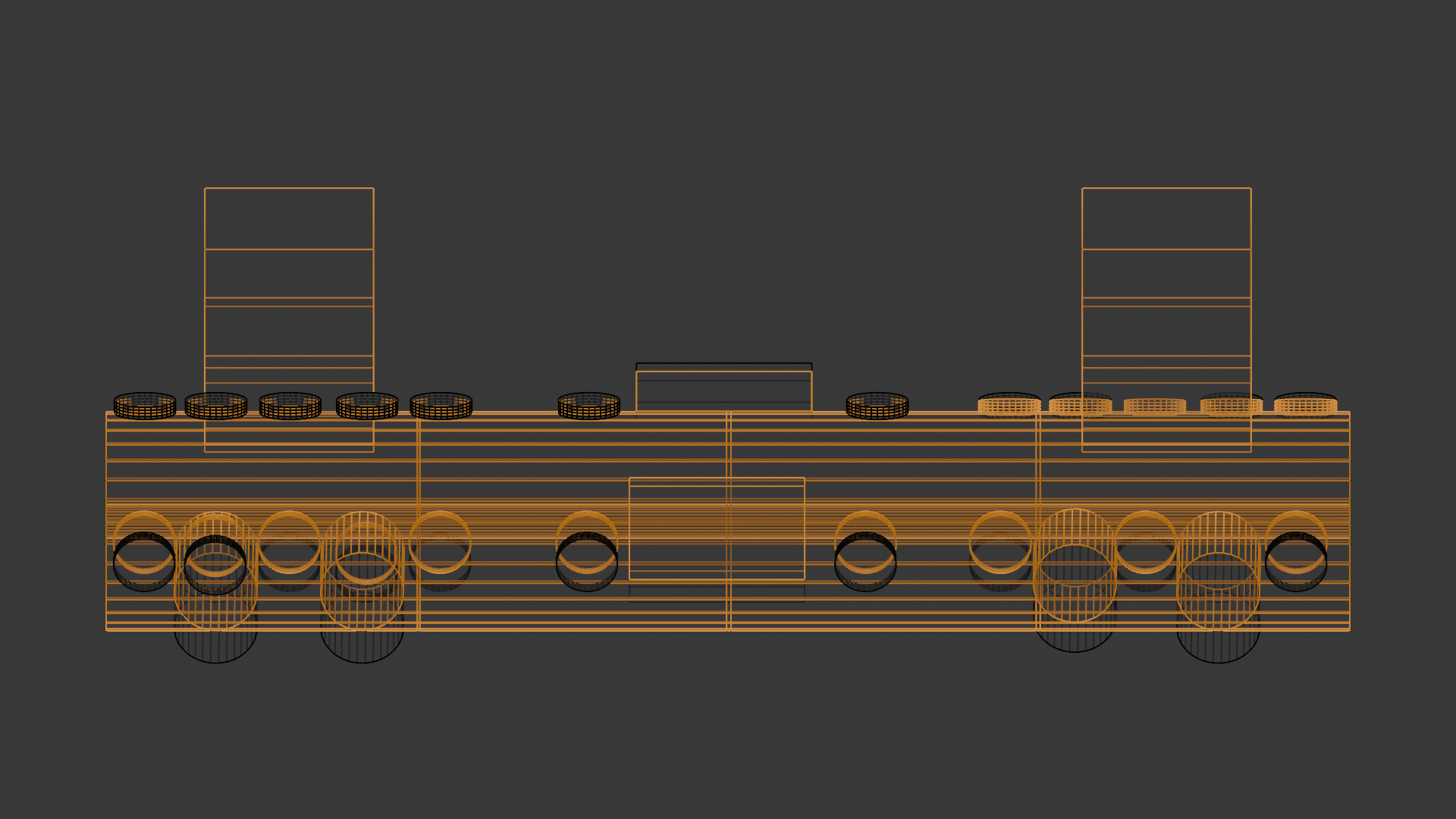}
        \caption{wireframe }
    \end{subfigure}
    \hfill
    \begin{subfigure}{0.3\textwidth}
        \includegraphics[width=\linewidth]{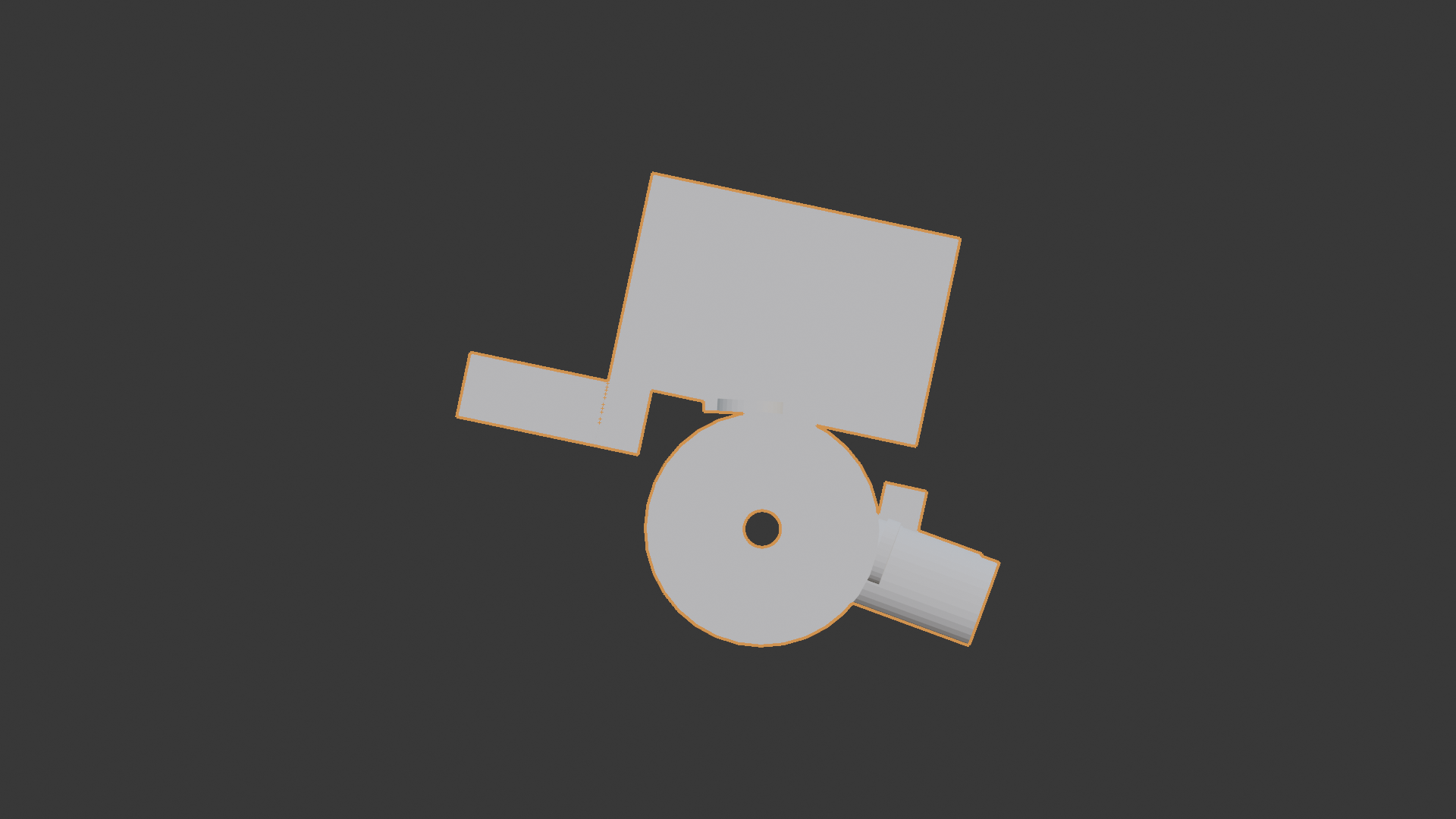}
        \caption{Side-View }
    \end{subfigure}
    \caption{RFQ-3D model }
\end{figure}

To accelerate negative hydrogen ions inside the RFQ a time-varying sinusoidal electric field was used according to the algorithm written for RFQ \cite{Wangler2008}. The formula used in the algorithm to generate the electric field is given by,
\begin{equation}
    E = E_0\sin(2\pi f t) 
\label{eq: rfq formula}
\end{equation}

\begin{itemize}

    \item\(E\):Instantaneous Electric field strength 
    \item\(E_0\):Amplitude of the Electric field  
    \item\(f\): Frequency of the oscillation

\end{itemize}

The particles are affected by this electric field as they enter the area of RFQ. The player can control the frequency using the joystick placed in the User Interface(UI).

\subsection{Chopper Line}

\raggedright
\justify
The 3D model for the Chopper Line is designed as the figure \label{fig: Chopperline_model}

\begin{figure}[H]
    \centering
    \begin{subfigure}{0.3\textwidth}
        \includegraphics[width=\linewidth]{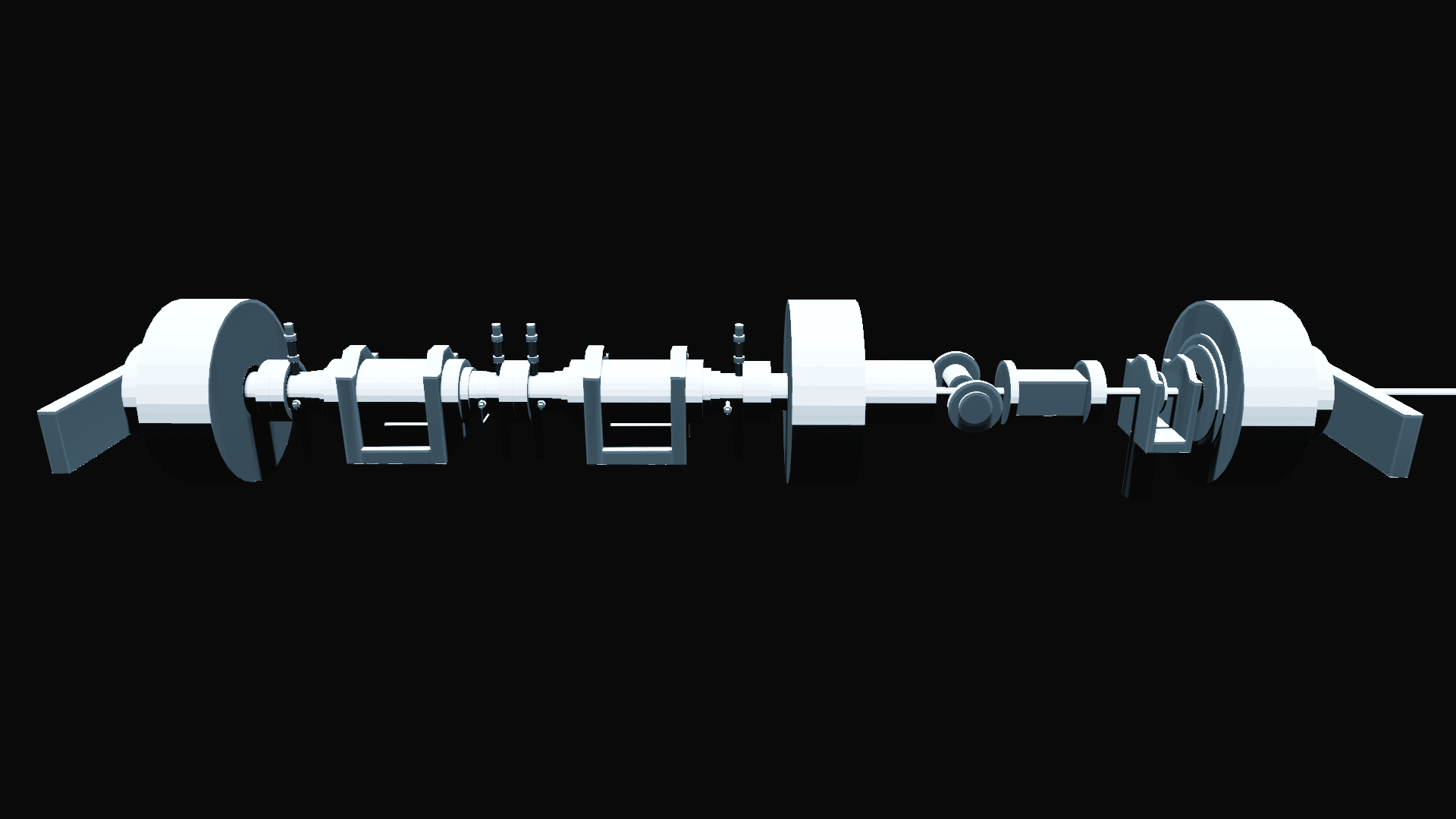}
        \caption{ Front view}
    \end{subfigure}
    \hfill
    \begin{subfigure}{0.3\textwidth}
        \includegraphics[width=\linewidth]{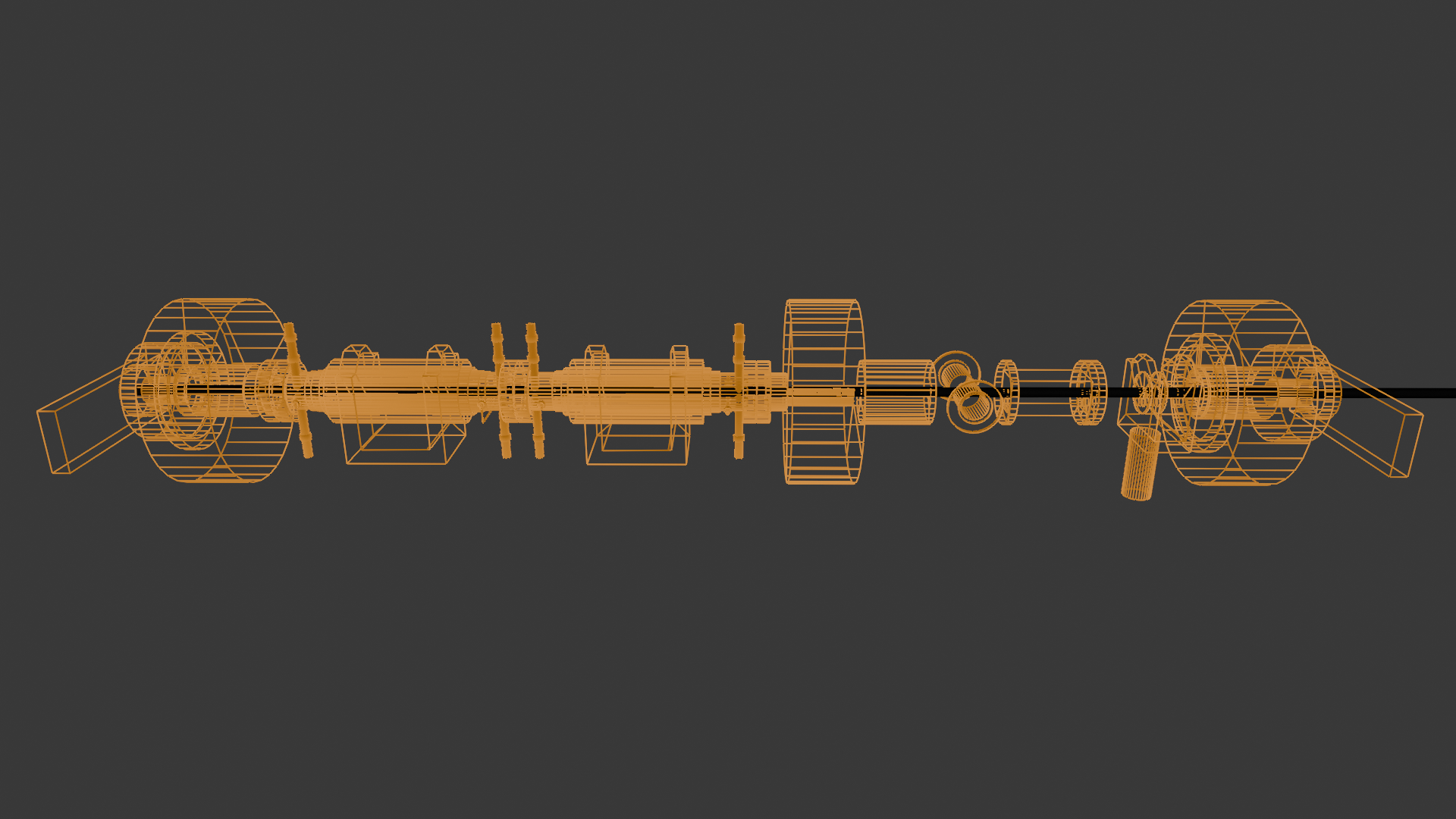}
        \caption{wireframe }
    \end{subfigure}
    \hfill
    \begin{subfigure}{0.3\textwidth}
        \includegraphics[width=\linewidth]{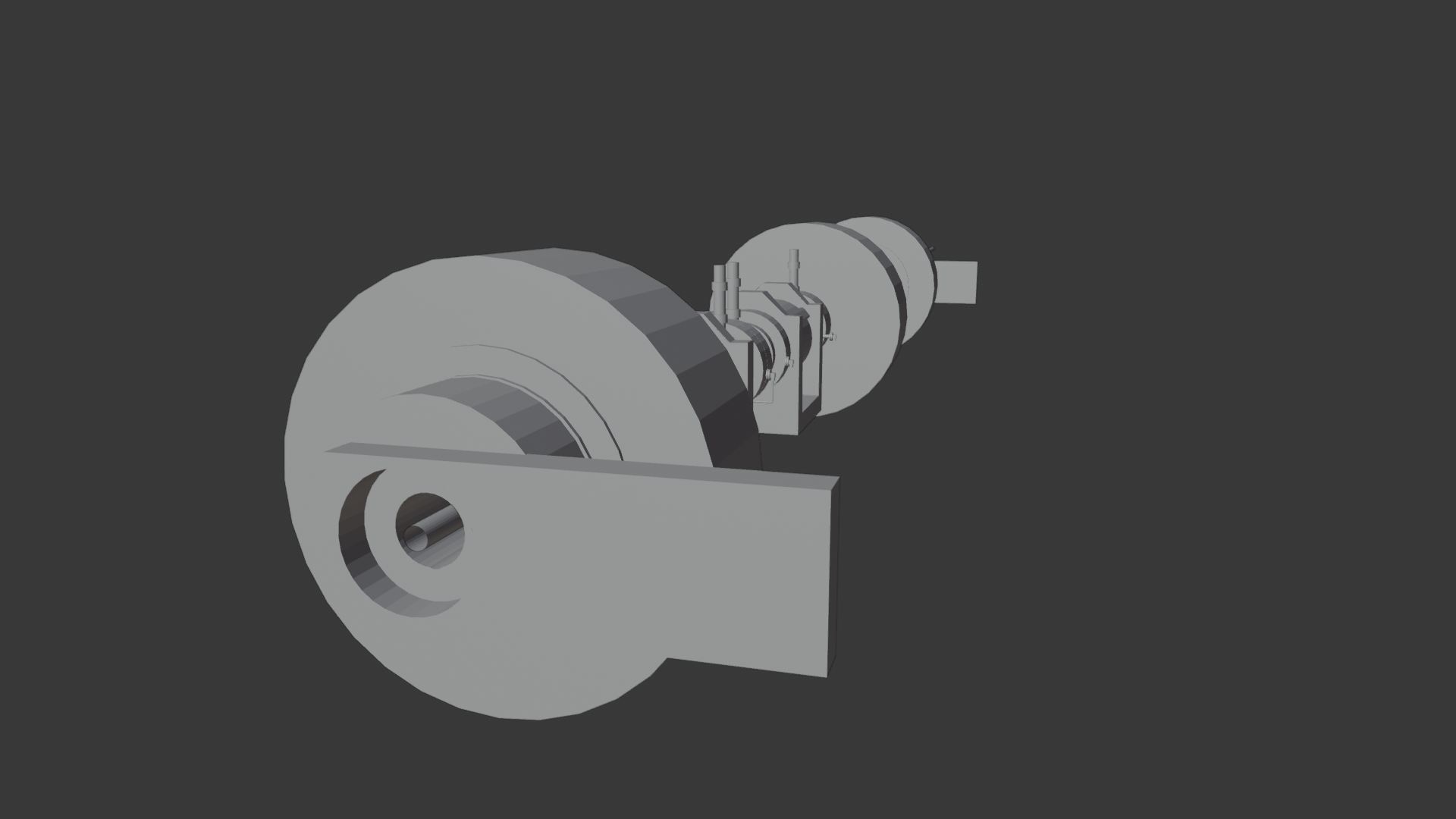}
        \caption{Side - View }
    \end{subfigure}
    \caption{Chopper Line-3D model }
\end{figure}

To match the beam with the frequency of the PSB  a Random chopping system was implemented, where a particle blocker is turned on and off according to an algorithm. The algorithm used for the chopper Line is as follows,
\newpage

\begin{equation}
Total\,chopping\,Time = \frac{ 1}{Chopping\,Frequency} 
\label{eq: algorithm   for chopping line}
\end{equation}

\begin{equation}
Beam\,On\,Time  = Total\,chopping\,Time  * Duty\,Cycle 
\label{eq: algorithm   for chopping line -2}
\end{equation}

\begin{equation}
Beam\,Off\,Time = Total\,chopping\,Time * (1 - Duty\,Cycle) 
\label{eq: algorithm   for chopping line -3 }
\end{equation}

\begin{itemize}

    \item\(Total\,chopping\,Time\):Time for a one chopping cycle 
    \item\(Duty\,Cycle\):Split factor  
    \item\( Beam\,On\,Time \): The time beam is allowed to flow within one chopping cycle.
    \item\( Beam\,off\,Time \): The time beam is not allowed to flow within one chopping cycle.
    
\end{itemize}

The chopping frequency and the split factor are controlled by the user and Total chopping time and beam on and off times are calculated using these formulas.
Then several active,inactive periods of the  blocker were calculated using the following formula,

\begin{equation}
    Ton_i =\frac {Beam\,On\,Time}{5} + i*\frac{ Beam\,On\,Time}{30} 
    \label{eq: algorithm   for chopping line -4 }
\end{equation}

\begin{equation}
    Toff_i =\frac {Beam\,Off\,Time}{5} + i*\frac{ Beam\,Off\,Time}{30} 
    \label{eq: algorithm   for chopping line -4 }
\end{equation}

\begin{itemize}

    \item\( Ton_i \) : $i^{th}$ Time period where beam is allowed to flow
    \item\( Toff_i \) : $i^{th}$ Time period where beam is not allowed to flow
   
\end{itemize}

For this algorithm, it was taken as one chopping cycle has four time periods that allow the beam to flow and another four time periods the beam is blocked.
The activation and deactivation processes of the blocker happen alternatively according to generated periods. After eight time periods, the next chopping cycle will begin.
the following diagram visually represents the time allocation according to the algorithm.

If time periods where beam is allowed to flow : $[t_a,t_b,t_c,t_d]$ and 
time periods where beam is not allowed to flow : $[t_e,t_f,t_g,t_h]$
\begin{center}
    \begin{tikzpicture}[node distance=1cm, every node/.style={draw, minimum width=1.5cm, minimum height=1cm, align=center}]
    \node (box1) at (0, 0) {ON};
    \node (box2) at (1.5, 0) {OFF};
    \node (box3) at (3, 0) {ON};
    \node (box4) at (4.5, 0) {OFF};
    \node (box5) at (6.0, 0) {ON};
    \node (box6) at (7.5, 0) {OFF};
    \node (box7) at (9.0, 0) {ON};
    \node (box8) at (10.5, 0) {OFF};
    
      
    \node at (box1.center) [above,yshift=0.7cm] {$t_a$}; 
    \node at (box2.center) [above,yshift=0.7cm] {$t_e$}; 
    \node at (box3.center) [above,yshift=0.7cm] {$t_b$}; 
    \node at (box4.center) [above,yshift=0.7cm] {$t_f$}; 
 
    \node at (box5.center) [above,yshift=0.7cm] {$t_c$}; 
    \node at (box6.center) [above,yshift=0.7cm] {$t_g$}; 
    \node at (box7.center) [above,yshift=0.7cm] {$t_d$}; 
    \node at (box8.center) [above,yshift=0.7cm] {$t_h$};

    \end{tikzpicture}
\end{center}

For a one-chopping cycle ,

$Beam\ on\ Time = t_a + t_b + t_c + t_d$ \\
      $\hspace*{17pt} Beam\ off\ Time = t_e + t_f + t_g + t_h$ \\
 $\hspace*{17pt} Total\ chopping \ Time = t_a + t_b + t_c + t_d + t_e + t_f + t_g + t_h$

\subsection{Drift  Tube  Linac (DTL)}

\raggedright
\justify
 The 3D model for the DTL was designed as the figure \cite{article},

 \begin{figure}[H]
     \centering
     \begin{subfigure}{0.3\textwidth}
         \includegraphics[width=\linewidth]{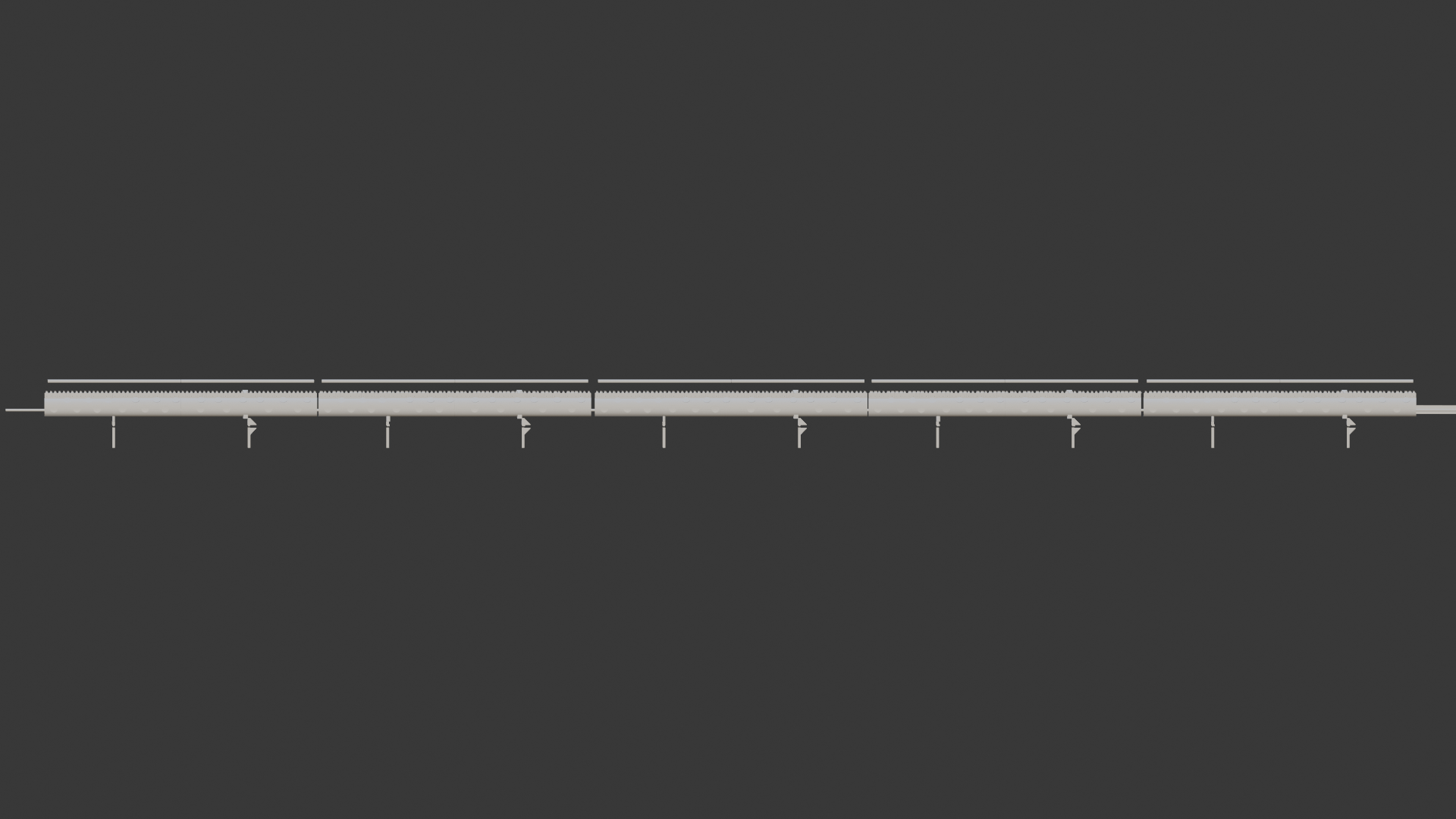}
         \caption{ Front View}
     \end{subfigure}
     \hfill
     \begin{subfigure}{0.3\textwidth}
         \includegraphics[width=\linewidth]{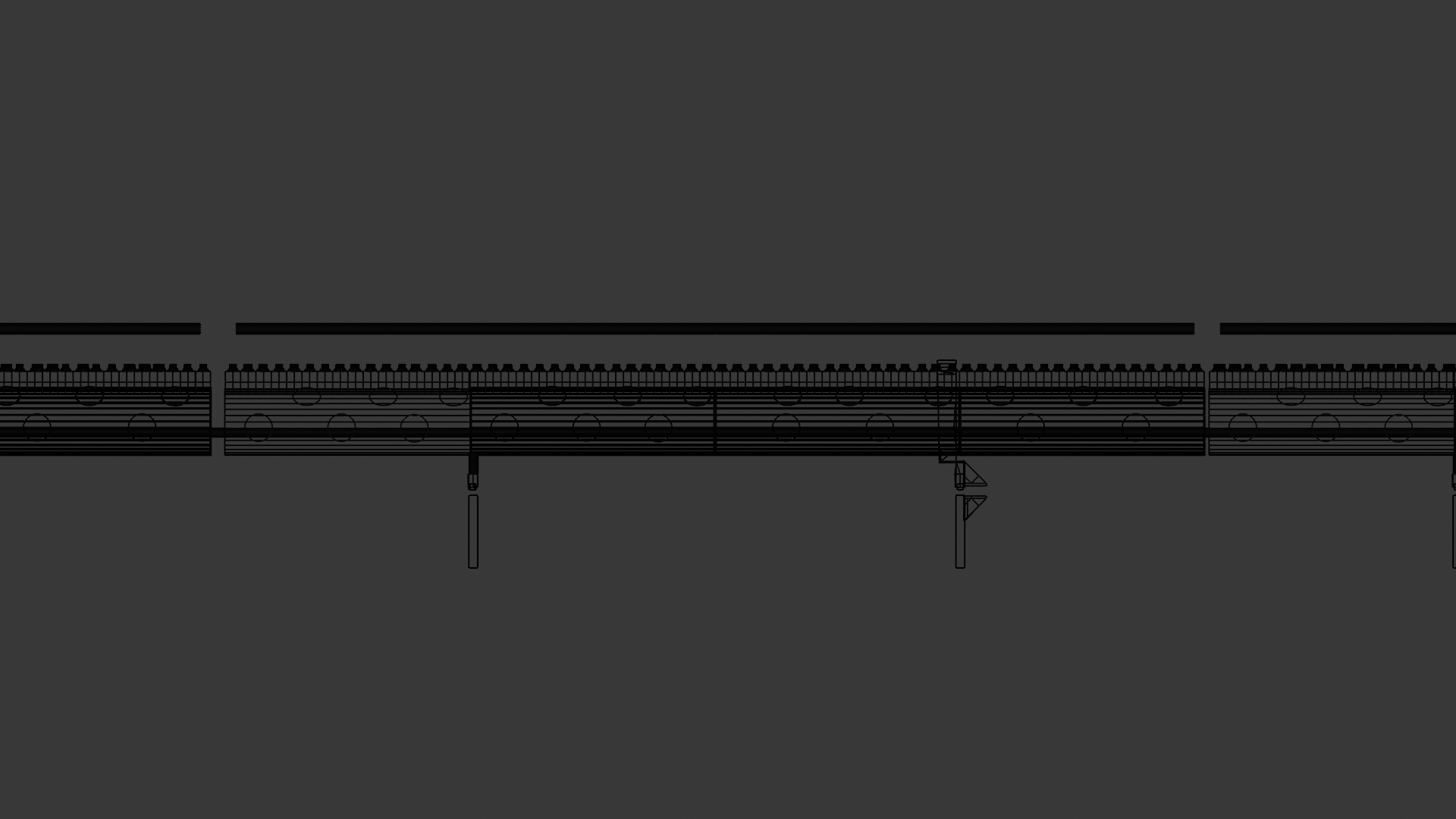}
         \caption{wireframe }
     \end{subfigure}
     \hfill
     \begin{subfigure}{0.3\textwidth}
         \includegraphics[width=\linewidth]{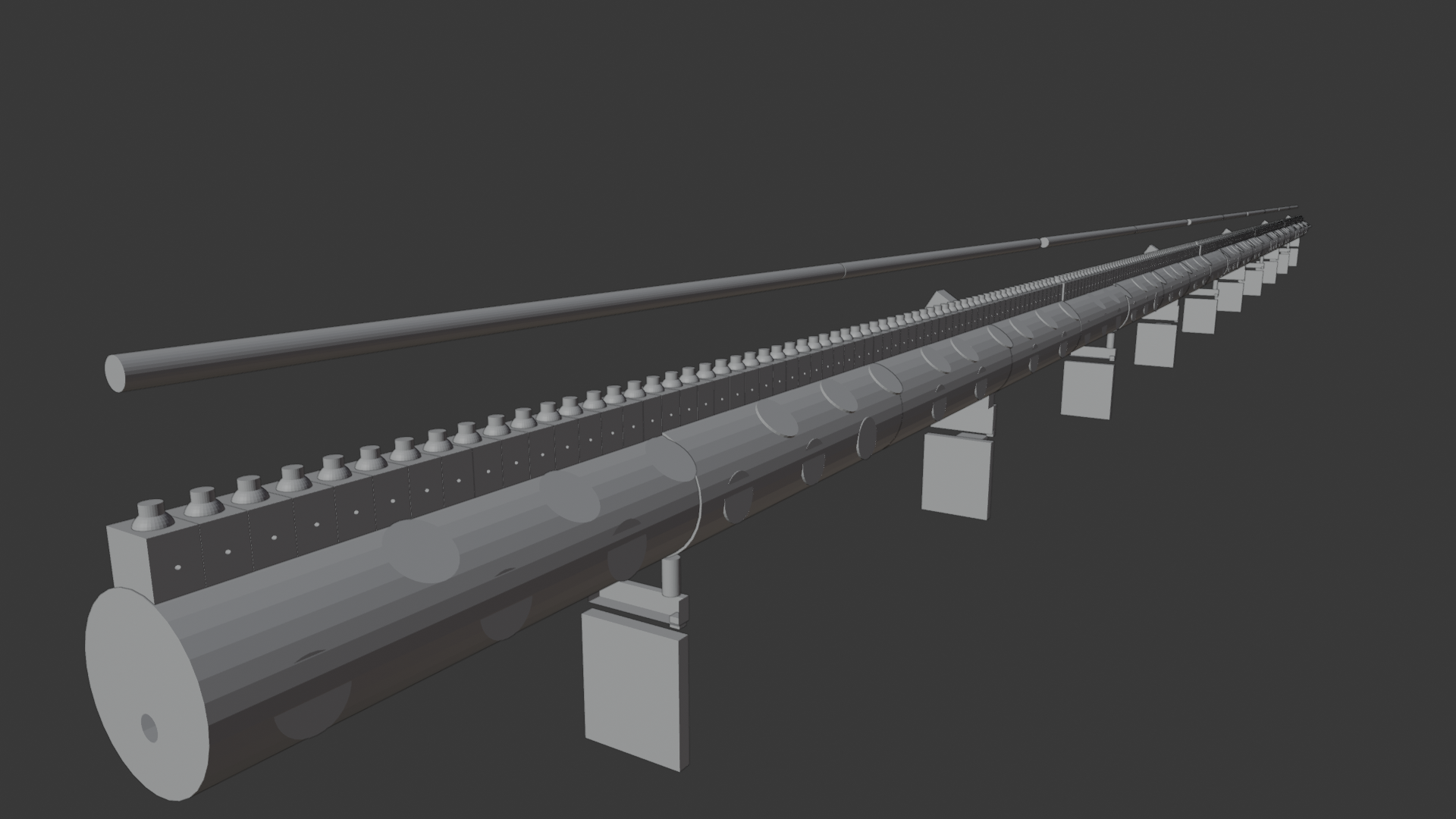}
         \caption{ Side View }
     \end{subfigure}
     \caption{DTL-3D model }
 \end{figure}
 
 To accelerate negative hydrogen ions inside the DTL same formula used in RFQ was used.
 \begin{equation}
     E = E_0\sin(2\pi f t) 
 \label{eq: algorithm  for ChopperLine}
 \end{equation}

 \begin{itemize}
 
     \item\(E\):Intantaneous Electric field strength 
     \item\(E_0\):Amplitude of the Electric field  
     \item\(f\): Frequency of the oscillation
 
 \end{itemize}

 In DTL, the algorithm was implemented as the electric field is only given when the particle is inside an area called a gap, which has an assigned value, gap length. 
 ten gaps were implemented inside DTL. Players can use the joystick (UI) to control the frequency of the electric field.

\subsection{Cell Coupled Drift  Tube  Linac (CCDTL)}

\raggedright
\justify
 The 3D model for the CCDTl was designed as the figure \label{fig: CCDTL}

 \begin{figure}[H]
     \centering
     \begin{subfigure}{0.3\textwidth}
         \includegraphics[width=\linewidth]{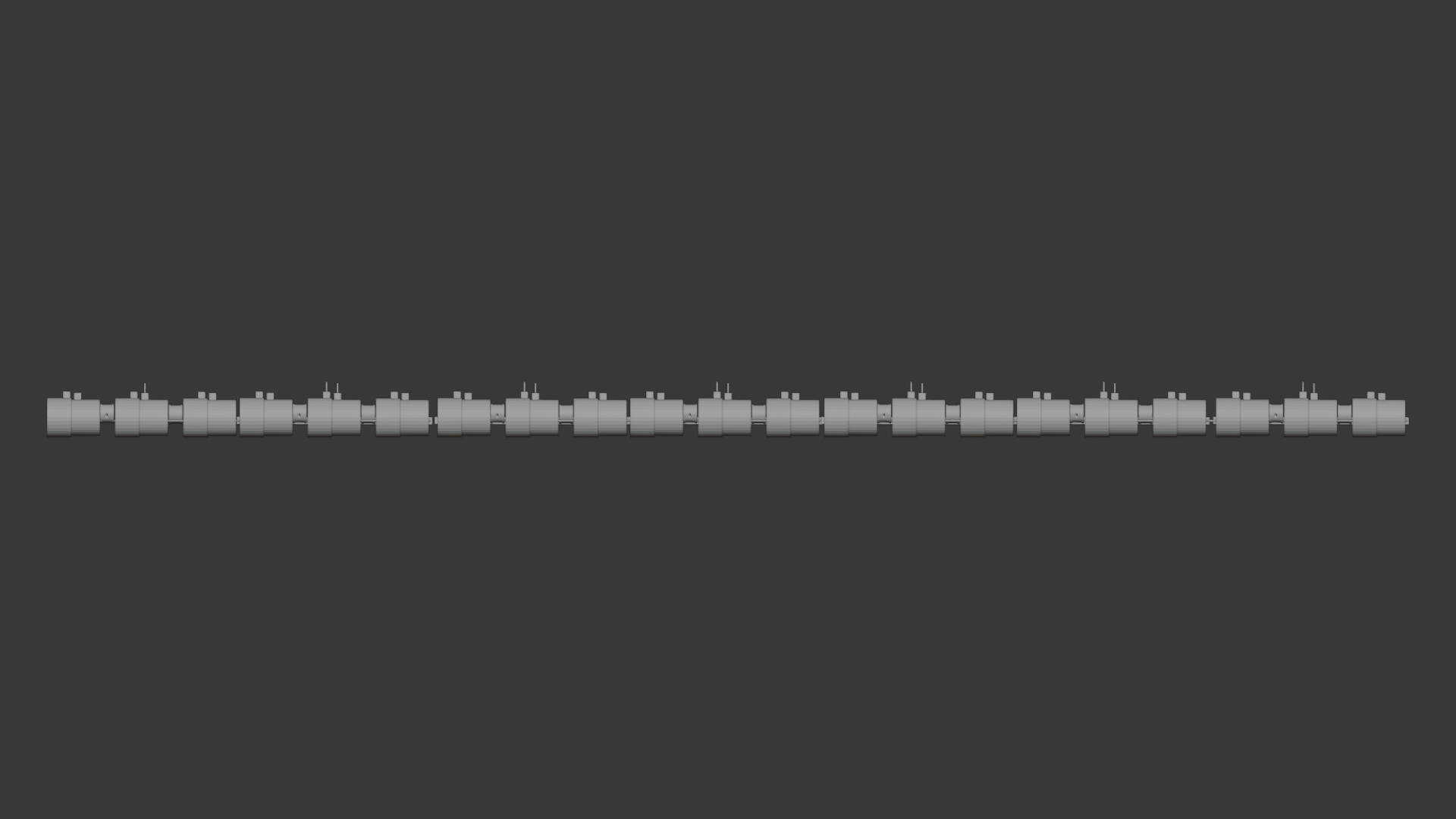}
         \caption{ Front-View}
     \end{subfigure}
     \hfill
     \begin{subfigure}{0.3\textwidth}
         \includegraphics[width=\linewidth]{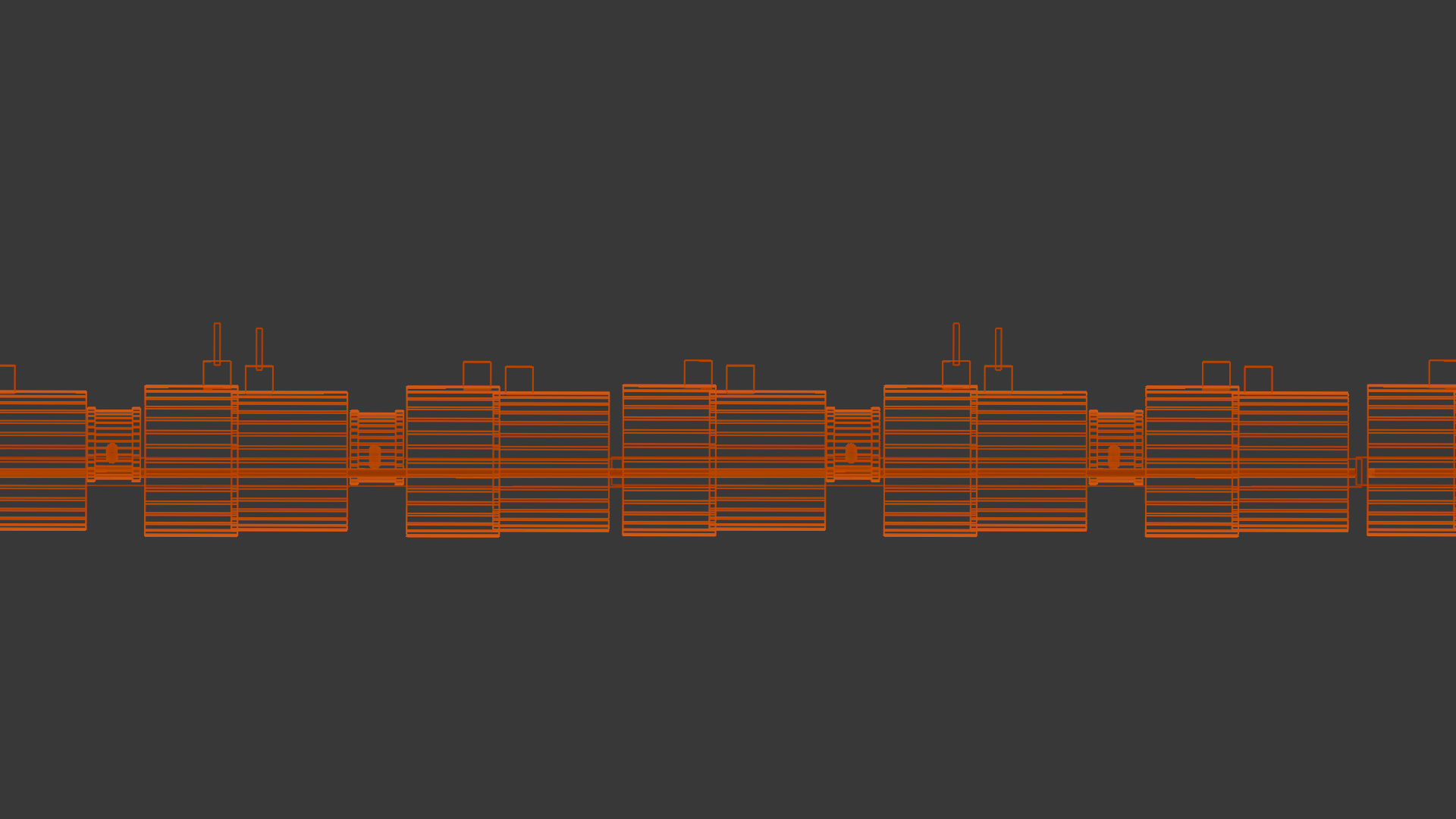}
         \caption{wireframe }
     \end{subfigure}
     \hfill
     \begin{subfigure}{0.3\textwidth}
         \includegraphics[width=\linewidth]{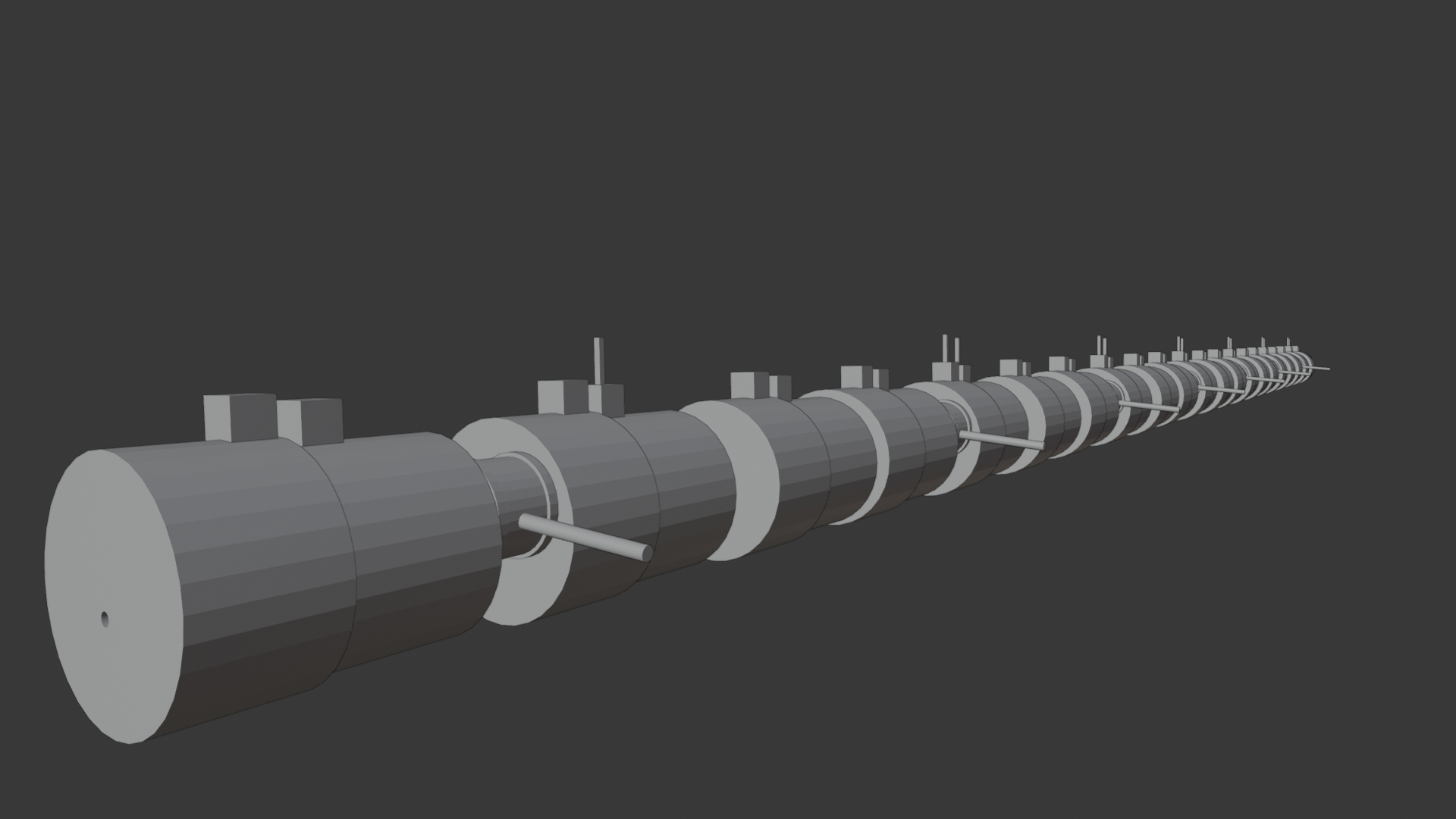}
         \caption{Side-View }
     \end{subfigure}
     \caption{CCDTL-3D model }
 \end{figure}
 
 To accelerate negative hydrogen ions inside the CCDTL following formula was used.
 \begin{equation}
     E=  E_0\sin(kx)cos(2\pi f t + \theta)
 \label{eq: algorithm  for ChopperLine}
 \end{equation}

 \begin{itemize}
 
     \item\(E\):Intantaneous Electric field strength 
     \item\(E_0\):Amplitude of the Electric field  
     \item\(f\): Frequency of the oscillation
     \item \(\theta\): Phase constant
 
 \end{itemize}

 In CCDTL, the electric field is only generated when the particle is inside a gap.
 Twenty-one gaps were implemented inside CCDTL. Players can use the joystick (UI) to control the frequency of the electric field.

\subsection{PI Mode Structures(PIMS)}

 The 3D model for the PIMS is designed as the figure \label{fig: PIMS}
 
 \begin{figure}[H]
     \centering
     \begin{subfigure}{0.3\textwidth}
         \includegraphics[width=\linewidth]{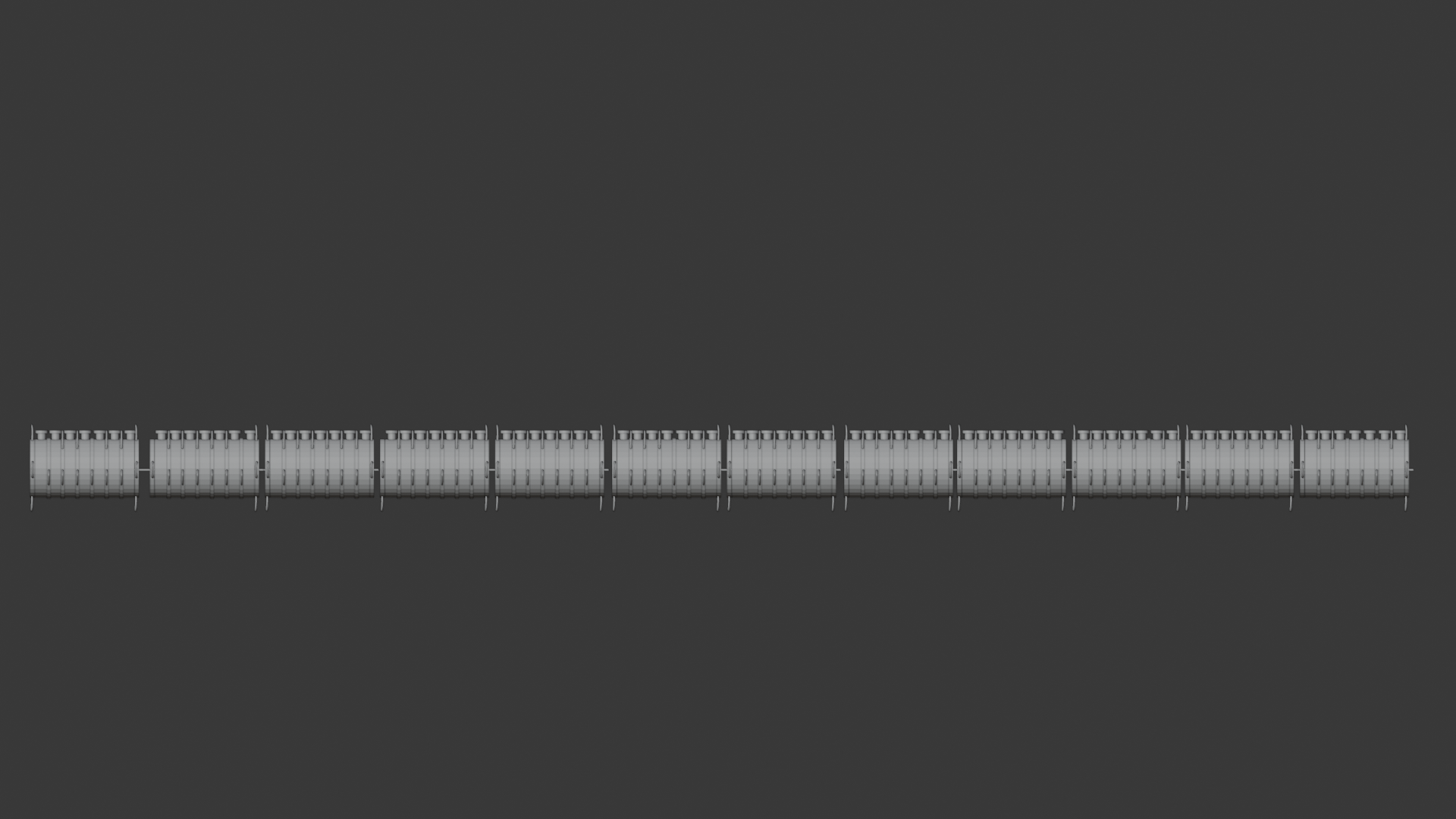}
         \caption{ Front view}
     \end{subfigure}
     \hfill
     \begin{subfigure}{0.3\textwidth}
         \includegraphics[width=\linewidth]{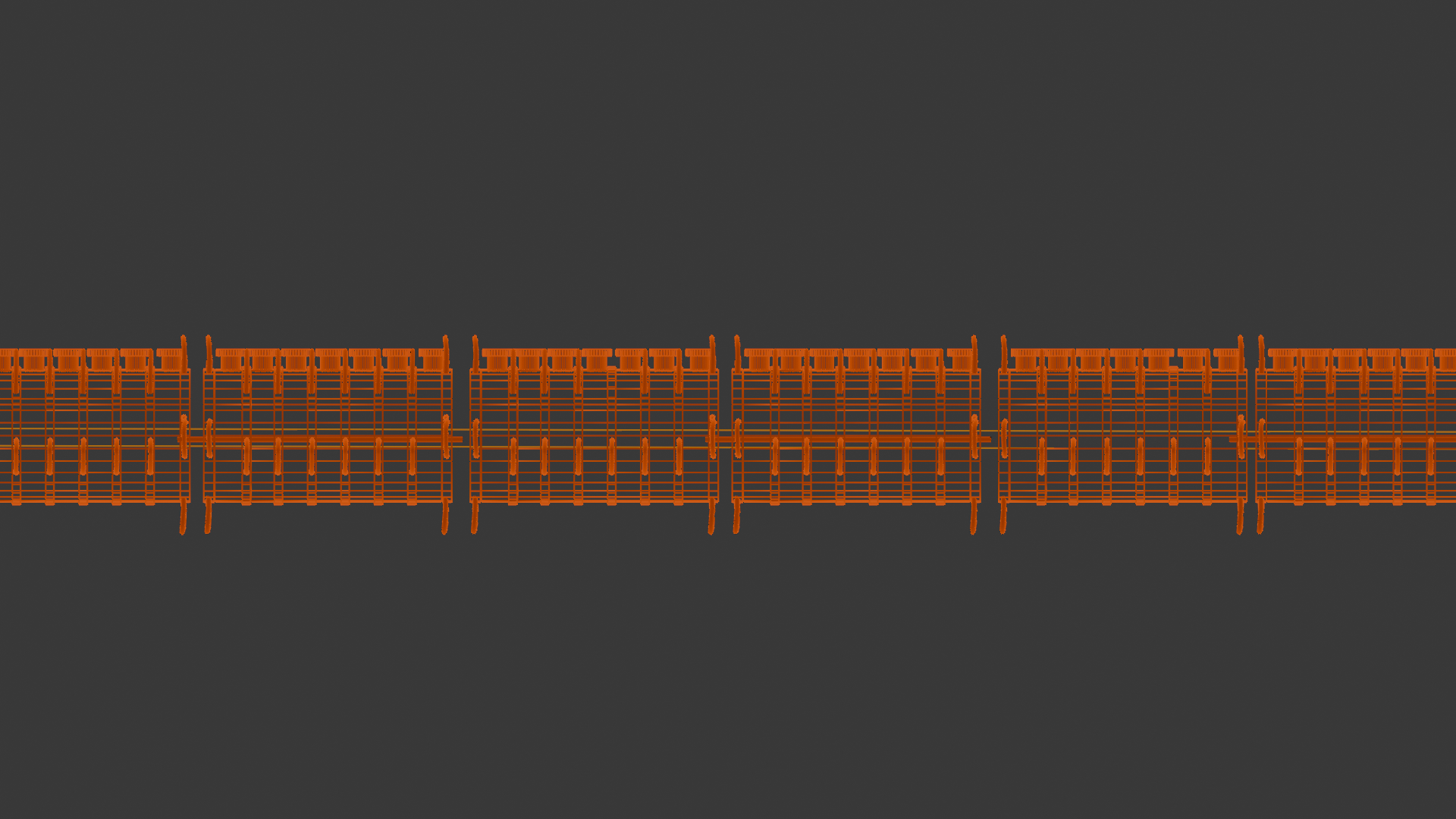}
         \caption{wireframe }
     \end{subfigure}
     \hfill
     \begin{subfigure}{0.3\textwidth}
         \includegraphics[width=\linewidth]{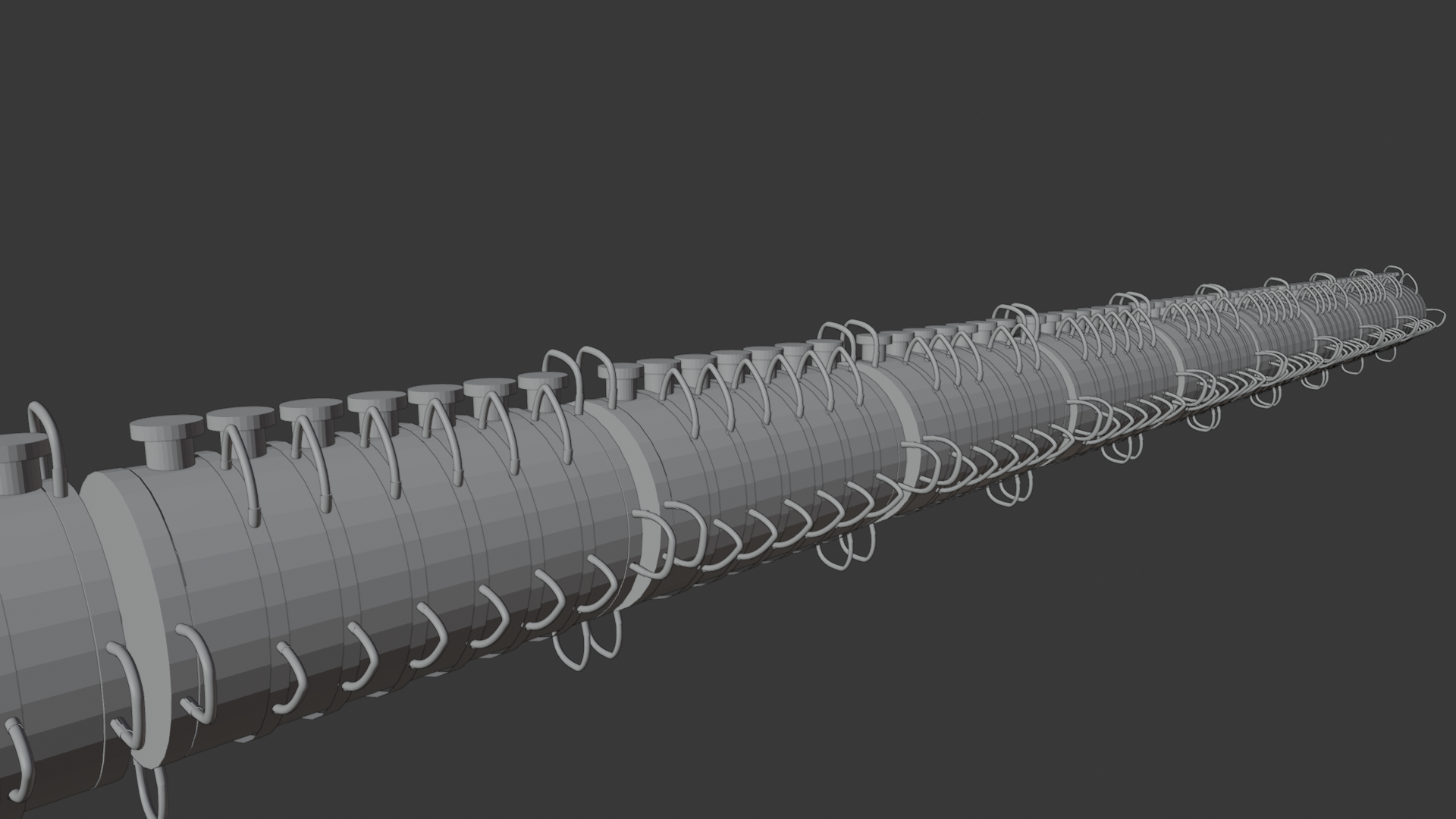}
         \caption{Side View }
     \end{subfigure}
     \caption{PIMS-3D model }
 \end{figure}
 
 The same waveform of the electric field as CCDTL was implemented for the PIMS \cite{Tirado2012}.
 The formula is given by,
 \newpage
 \begin{equation}
     E=  E_0\sin(kx)cos(2\pi f t + \theta)
 \label{eq: algorithm  for ChopperLine}
 \end{equation}

 \begin{itemize}
 
     \item\(E\):Intantaneous Electric field strength 
     \item\(E_0\):Amplitude of the Electric field  
     \item\(f\): Frequency of the oscillation
     \item \(\theta\): Phase constant
 
 \end{itemize}

 In PIMS, the algorithm was designed as the electric field can be generated throughout the entire machine. Players can use the joystick (UI) to control the frequency of the electric field.

In LINAC-4 simulation model, movement of the $H^-$ particle was simulated using the above-discussed algorithm. 
\raggedright
\justify

The main aim of the study was to design an interactive interface/game to simulate negative H ion acceleration in LINAC-4.The accelerating motion of negative hydrogen ions inside LINAC-4 was simulated using varying electric fields in different waveforms, where the player controls the frequency. The real-time velocity vs time graph of the negative hydrogen ion in LINAC-4  was added to visualize the velocity change effectively. In this simulation study, for each machine, several types of challenges were raised and simulation was done with a lot of limitations. The workflow mechanisms of RFQ, DTL,CCDTL, and PI mode structures consist of very
complex sub-sections that are hard to simulate. LINAC-4 consists of two main
workflows: accelerating and focusing the ion beam. Simulating both accelerating
and focusing mechanisms for every sub-section of LINAC-4 is complex and time consuming. In this simulation, only the accelerating function was considered. Each subsection of LINAC-4 has a bunch of pre workflows not directly affect the beam but must needed for final operation to be happen. Simulating all these pre-workflows is a very hard task. So only the final effect on the beam by each section was considered and simulated. Also velocity of the negative hydrogen ion was taken according to a scale: of 1:$10^6$, for simplicity and to avoid complex calculations.

\bibliographystyle{jhep}
\bibliography{ref}

\end{document}